\documentclass[preprint,12pt]{elsarticle}

\usepackage{graphicx}
\usepackage{color}
\usepackage[colorlinks,bookmarks=false,citecolor=blue,linkcolor=red,urlcolor=blue]{hyperref}

\usepackage{amssymb,amsbsy}
\usepackage{amsmath}

\newcommand{\op}[1]{%
    \fontdimen12\textfont3=2pt\fontdimen12\scriptfont3=1.4pt%
    \!\null\mathop{\vphantom{#1}\smash{#1}}\limits_{\sim}\null\!}

\newcommand{\ddt}{\frac{d}{dt}\;}
\newcommand{\fmref}[1]{(\protect\ref{#1})}
\newcommand{\Operator}[1]{%
    \fontdimen12\textfont3=2pt\fontdimen12\scriptfont3=1.4pt%
    \!\null\mathop{\protect\vphantom{#1}\smash{#1}}\limits_{\sim}\null\!}

\journal{J. Magn. Magn. Mater.}

\begin{document}

\begin{frontmatter}



\title{Investigation of thermalization in giant-spin models by different Lindblad schemes}


\author{Christian Beckmann\fnref{bi}}
\author{J{\"u}rgen Schnack\corref{cor1}\fnref{bi}}
\ead{jschnack@uni-bielefeld.de}
\cortext[cor1]{corresponding author}
\address[bi]{Dept. of Physics, Bielefeld University, P.O. box
  100131, D-33501 Bielefeld, Germany}

\begin{abstract}
The theoretical understanding of time-dependence in magnetic
quantum systems is of great importance in particular for cases
where a unitary time evolution is accompanied by relaxation
processes. A key example is given by the dynamics of
single-molecule magnets where quantum tunneling of the
magnetization competes with thermal relaxation over the
anisotropy barrier. In this article we investigate how good a
Lindblad approach describes the relaxation in giant spin models
and how the result depends on the employed operator that
transmits the action of the thermal bath.
\end{abstract}


\end{frontmatter}


\section{Introduction}
\label{sec-1}

Single-molecule magnets (SMM) show two interesting phenomena:
slow relaxation and quantum tunneling of the magnetization
\cite{SGC:Nat93,TLB:Nature96,WeS:Science99}. Very often both
processes are modeled independently of each other. Relaxation
is accounted for by rate equations, see
e.g. Refs.~\cite{CWM:PRL00,CGS:PRB05,SCL:PRL05,CGS:PRL08,GBC:PRB16},
whereas quantum tunneling is described by a unitary time
evolution \cite{CGS:PRB05}, 
e.g. in terms of the von~Neumann equation. Only a few approaches
have been undertaken in order obtain a combined description.

In this article we investigate a master-equation approach that
rests on the use of Lindblad terms
\cite{Lin:CMP76,RMS:PRB97,TRJ:PLA98,MSD:PRL98,SMD:PRB99,SaM:JPSJ01,NaM:JPSJ01,KNT:P09,CGT:PRA14}.
Although such an approach lacks memory effects
\cite{TRJ:CP98,TRJ:PLA98} it constitutes a minimal feasible
description of a time evolution that combines coherent and
incoherent parts. Besides application for magnetization dynamics
such a description is of paramount importance for the
investigation of quantum computing schemes and their robustness
\cite{CGT:PRA14}.

The Lindblad terms in the master equation usually summarize the
effects of the various relaxation processes \cite{LvS:CSR15}
in terms of transition operators acting between the eigenstates 
of the investigated quantum spin system. Although it might in
principle be possible to derive such terms from basic principles
\cite{SCL:PRL05,CRP:NM16}, the employed functional form of these terms
leaves room to tweak unknown parameters \cite{KNT:P09}. In the
following we introduce some of the common approaches and
investigate how the resulting magnetization dynamics depends on
the parameterization. Our impression is that this dependence is
in general non-negligible and in particular for the evaluation
of the ac magnetization rather strong, so that results from such
simulations have to be interpreted with great care.

The article is organized as follows. In section~\ref{sec-2} we
introduce the model. Sections~\ref{sec-3} and \ref{sec-4} deal
with dc and ac magnetization, respectively. Our results are
summarized in section~\ref{sec-5}.

\section{Model and numerical procedures}
\label{sec-2}

For the investigations throughout this article we employ the
following giant spin Hamiltonian \cite{CGS:JMMM99}
\begin{eqnarray}
\label{E-2-1}
\op{H}
&=&
D\;
\op{S}_z^2
+
E\;
\left(
\op{S}_x^2 - \op{S}_y^2
\right)
+
g\, \mu_B\, \vec{B}(t)\cdot
\op{\vec{S}}
\ ,
\end{eqnarray}
with specific values of $D=-1\text{~K}$, $E=0.1\text{~K}$ and
$g=2$ used in the numerical simulations. Such Hamiltonians
describe SMMs with an easy axis anisotropy of strength $D<0$.

For the time evolution of the density matrix we employ \cite{KNT:P09}
\begin{eqnarray}
\label{E-2-2} 
\ddt\op{\rho}\left(t\right) 
&=&
-i\left[\op{H},\op{\rho}\left(t\right)\right]
-\lambda\left(\left[\op{X},\op{R}\op{\rho}\left(t\right)\right]
+\left[\op{X},\op{R}\op{\rho}\left(t\right)\right]^{\dagger}\right)
\ ,
\end{eqnarray}
with
\begin{eqnarray}
\label{E-2-3} 
\left\langle k\right|\op{R}\left|n\right\rangle  
&=&
\frac{I\left(E_{k}-E_{n}\right)-I\left(E_{n}-E_{k}\right)}{e^{\beta\left(E_{k}-E_{n}\right)}-1}\left\langle
k\right|\op{X}\left|n\right\rangle  
\ ,
\end{eqnarray}
where $\left|k\right\rangle $ and $\left|n\right\rangle $ are eigenvectors
of the Hamiltonian with corresponding eigenvalues $E_{k}$ and
$E_{n}$. In such an approach it is implicitly assumed that the
coupling to the heat bath is realized by phonons
\cite{SMD:PRB99}, whose spectral 
density is denoted by $I\left(\omega\right)$
\begin{eqnarray}
I\left(E_{k}-E_{n}\right) & = & I_{0}\cdot\left(E_{k}-E_{n}\right)\label{E-2-4}
\end{eqnarray}
with $I_{0}=1$. The related
transition operator $\op{X}$ mediates the action of the heat
bath onto the giant spin of \fmref{E-2-1}. We investigate
several typical choices for this operator. For one choice of transition
operators we use those suggested in \cite{SMD:PRB99}:
\begin{eqnarray}
\op{X}_{1} & = & \op{S}_{x}\ ,\label{E-3-1}\\
\op{X}_{2} & = & \frac{1}{2}\left(\op{S}_{x}+\op{S}_{z}\right)
\ .\label{E-3-2}
\end{eqnarray}
These two operators generate transitions among eigenstates of
Hamiltonian \fmref{E-2-1} with $\Delta m = 0, \pm 1$. We
investigate as well another operator 
\begin{eqnarray}
\op{X}_{3} & = & \frac{4}{5}\op{S}_{x}+\frac{1}{5}\op{S}_{x}^{2}
\ ,\label{E-3-3}
\end{eqnarray}
that via its $\op{S}_{x}^{2}$ contribution effectively takes
also two-phonon processes with $\Delta m = \pm 2$ into account.

The influence of a spin bath, e.g. nuclear spins in the sample,
is not taken into account by this approach \cite{PrS:RPP00}.

\section{Magnetization dynamics: tunneling and hysteresis}
\label{sec-3}

In order to investigate the magnetization dynamics the magnetic
field in $z$-direction is swept from $-2$ to
$+2\text{~T}$ with a sweep-rate of $0.625\text{~T/ns}$.
For the simulations the sweep rate has to
assume such high values since otherwise simulations covering the complete
magnetization process are not feasible \cite{SMD:PRB99}. This
problem is caused by 
the intrinsic energy scales (GHz frequencies) and it is common to many
problems in simulation science. The gaps at the avoided level
crossings are adapted accordingly by choosing a constant field of 
$B_{x}=0.2\text{~T}$ in $x$-direction.

\begin{figure}[ht!]
	\centering
	\includegraphics[width=0.90\textwidth]{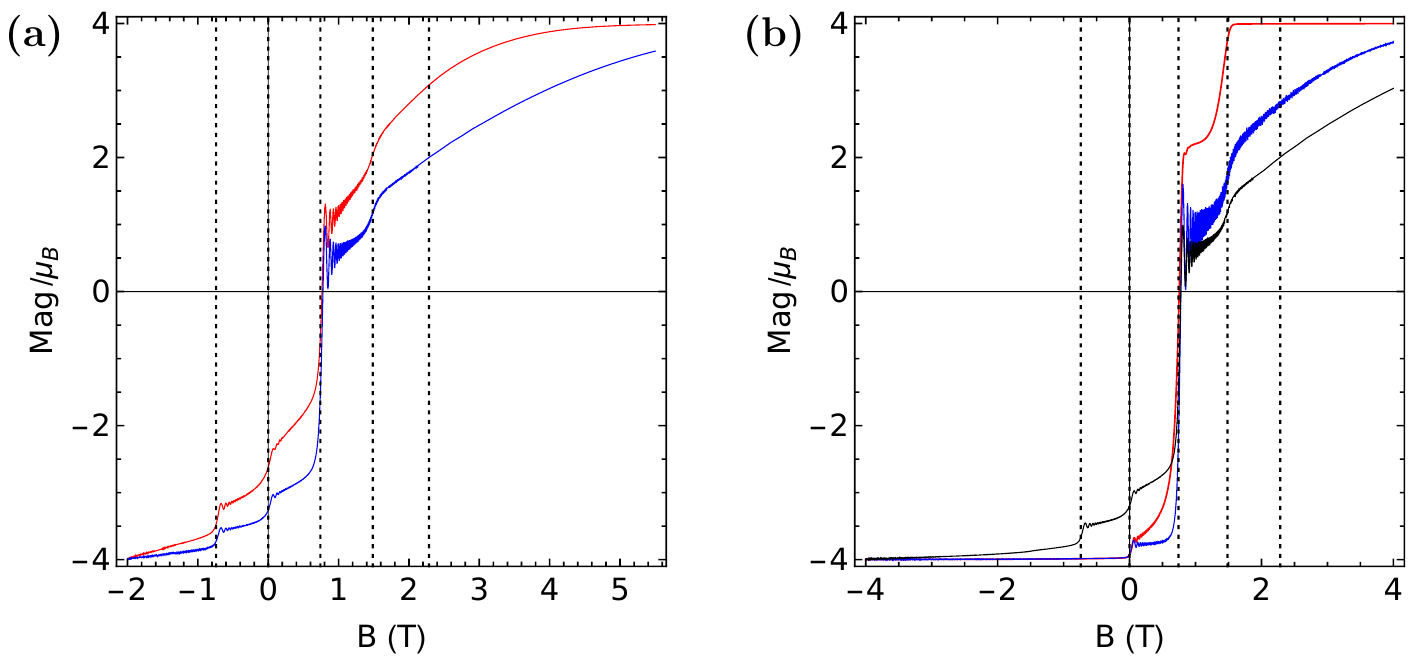}
	\caption{Magnetization in $z$-direction: The vertical dashed lines correspond
to magnetic field strengths where resonant tunneling is possible.
(a) $T=2$~K and $\lambda=10^{-4}$ with $\Operator{X}_{1}$ (red)
and $\Operator{X}_{2}$ (blue). (b) Magnetization curve with $\Operator{X}_{2}$
and $T=0.01$~K, $\lambda=10^{-2}$ (red), $T=0.01$~K, $\lambda=10^{-4}$
(blue), $T=2$~K, $\lambda=10^{-4}$ (black).} 
	\label{jmmm-beckmann-f-1}
\end{figure}

\begin{figure}[ht!]
	\centering
	\includegraphics[width=0.90\textwidth]{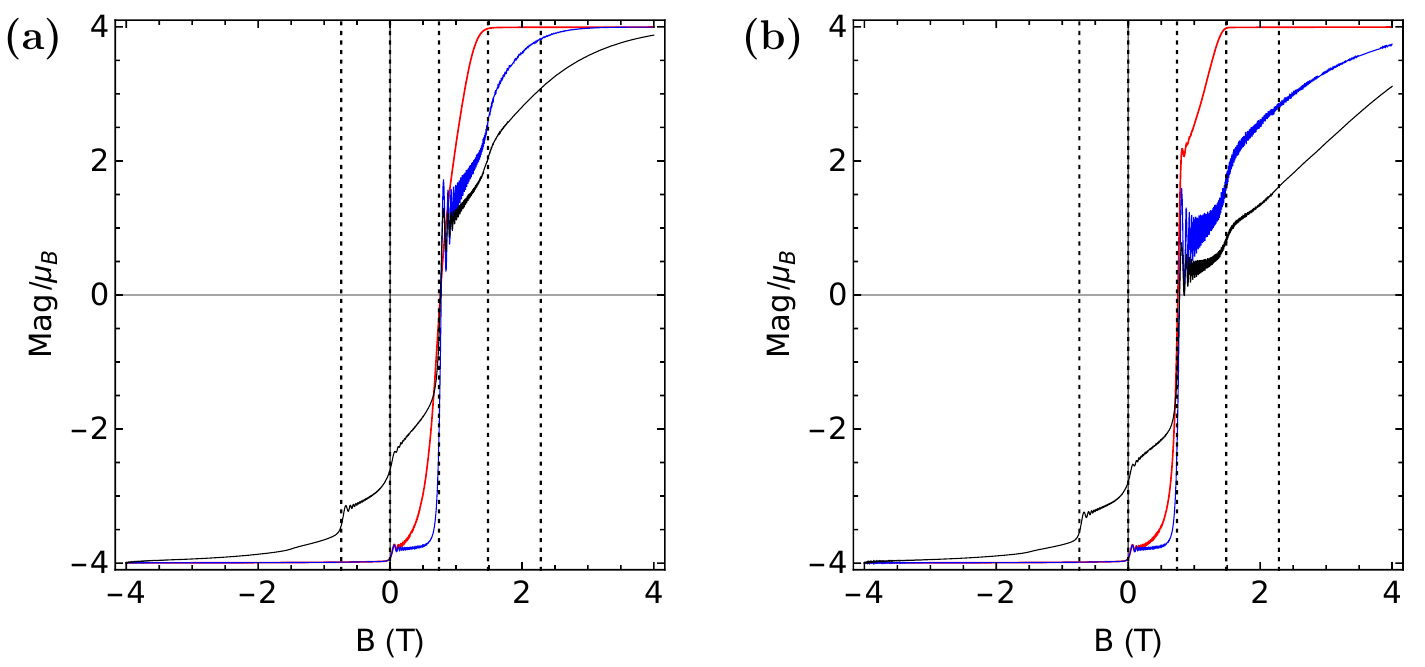}
	\caption{Magnetization in $z$-direction: The vertical dashed lines correspond
to magnetic field strengths where resonant tunneling is possible.
(a) Magnetization curve with $\Operator{X}_{1}$ and $T=0.01$~K,
$\lambda=10^{-2}$ (red), $T=0.01$~K, $\lambda=10^{-4}$ (blue),
$T=2$~K, $\lambda=10^{-4}$ (black). (b) Magnetization curve with
$\Operator{X}_{3}$ and $T=0.01$~K, $\lambda=10^{-2}$ (red), $T=0.01$~K,
$\lambda=10^{-4}$ (blue), $T=2$~K, $\lambda=10^{-4}$ (black).} 
	\label{jmmm-beckmann-f-2}
\end{figure}

Figure \ref{jmmm-beckmann-f-1} shows the step-wise behavior of the magnetization
in $z$-direction for the first two choices of $\op{X}$. All
magnetization steps appear at magnetic
field strengths where resonant tunneling is possible (as indicated
by the vertical dashed lines in Figure~\ref{jmmm-beckmann-f-1}). As already
predicted in \cite{SMD:PRB99} one can see in figure
\ref{jmmm-beckmann-f-2}~(a) that the relaxation with
$\op{X}=\op{X_{1}}$ is more efficient than the one with $\op{X}=\op{X}_{2}$
in the sense that the magnetization reaches saturation much
faster. This is very likely 
due to the factor $\frac{1}{2}$ in $\op{X}_{2}$, which reduces
the transition rates induced by $\op{S}_{x}$. Beside this
difference in efficiency both curves look qualitatively similar.

Figure \ref{jmmm-beckmann-f-1}~(b) and figure
\ref{jmmm-beckmann-f-2}~(a) \& (b) 
show magnetization curves for all three transition operators at different
temperatures and with different coupling constants $\lambda$. In
every case the curves calculated at $T=2$~K (black curves in figs.
\ref{jmmm-beckmann-f-1}~(b) and \ref{jmmm-beckmann-f-2}) show steps, that can only
occur because the involved level-crossings are already thermally
occupied. For $\op{X}=\op{X}_{1}$ and $\op{X}_{3}$ these steps have
qualitatively nearly the same step size, only in the curve with $\op{X}=\op{X}_{2}$
they are smaller, most likely due to the factor $\frac{1}{2}$ in
$\op{X}_{2}$. The thermal relaxation after the last step is also
different. Qualitatively the efficiency is increasing with the amount
of $\op{S}_{x}$ in the relaxation operator. The behavior at $T=0.01$~K
and $\lambda=10^{-2}$ (red curves in
figs. \ref{jmmm-beckmann-f-1}~(b) and \ref{jmmm-beckmann-f-2})
also shows differences in the efficiency. The curve 
with $\op{X}=\op{X}_{1}$ is again the most efficient one. Only two
steps are needed to end up at the saturation magnetization. Both other
curves show a third step. The visibility of this step increases with
the decreasing amount of $\op{S}_{x}$ in the transition
operator. 

As expected, a stronger coupling to the bath, i.e. a larger
parameter $\lambda$ leads to a quicker relaxation with reduced
quantum oscillations (compare red and blue curves in figures
\ref{jmmm-beckmann-f-1}~(b) and \ref{jmmm-beckmann-f-2}).

\section{Magnetization dynamics: AC susceptibility and
  relaxation times}
\label{sec-4}

AC susceptometry is a powerful experimental tool to get access
to the relaxation processes. Theoretically ac susceptibilities
are complicated non-equilibrium quantities since they involve the
influence of, and thus the coupling to the bath. It is thus
expected that properties of the bath and the coupling will in
general influence ac measurements \cite{PrS:RPP00}.

In our simulations we apply a magnetic ac field of the form
\begin{eqnarray}
B_{z}\left(t\right) & = & B_{0}\cdot\cos\left(\omega t\right)
\end{eqnarray}
in $z$-direction. It is important that the amplitude $B_{0}$ is small
enough, so that the magnetization of the system does not reach its
saturation value. We choose $B_{0}=0.001\text{ T}$. In the steady
state the magnetization of the system is given by \cite{GSV:2006}:
\begin{eqnarray}
M\left(t\right) & = & \left(\chi^{'}\cos\left(\omega t\right)+\chi^{''}\sin\left(\omega t\right)\right)B_{0}\:,
\end{eqnarray}
where $\chi^{'}$ and $\chi^{''}$ are the real and imaginary part
of the magnetic susceptibility. Both can be calculated from $M\left(t\right)$
via integration over one period. The relaxation time $\tau$ can
then be found as 
\begin{eqnarray}
\tau & = & \frac{1}{\omega_{\text{max}}}
\end{eqnarray}
with $\omega_{\text{max}}$ the frequency of the maximum of $\chi^{''}$.
The great advantage of this method is that it is possible to calculate
the relaxation times in the temperature regime where quantum tunneling
is the dominant process and the relaxation time becomes temperature
independent.

In our simulations of the ac susceptibility we set the coupling
to the phonon heat bath $\lambda=1$. 
We want to stress here, that the observed behavior is also present
with much smaller coupling constants. Smaller coupling constants only
lead to a shift of the temperature at which quantum tunneling becomes
the dominant process. 
\begin{figure}[ht!]
	\centering
	\includegraphics[width=0.90\textwidth]{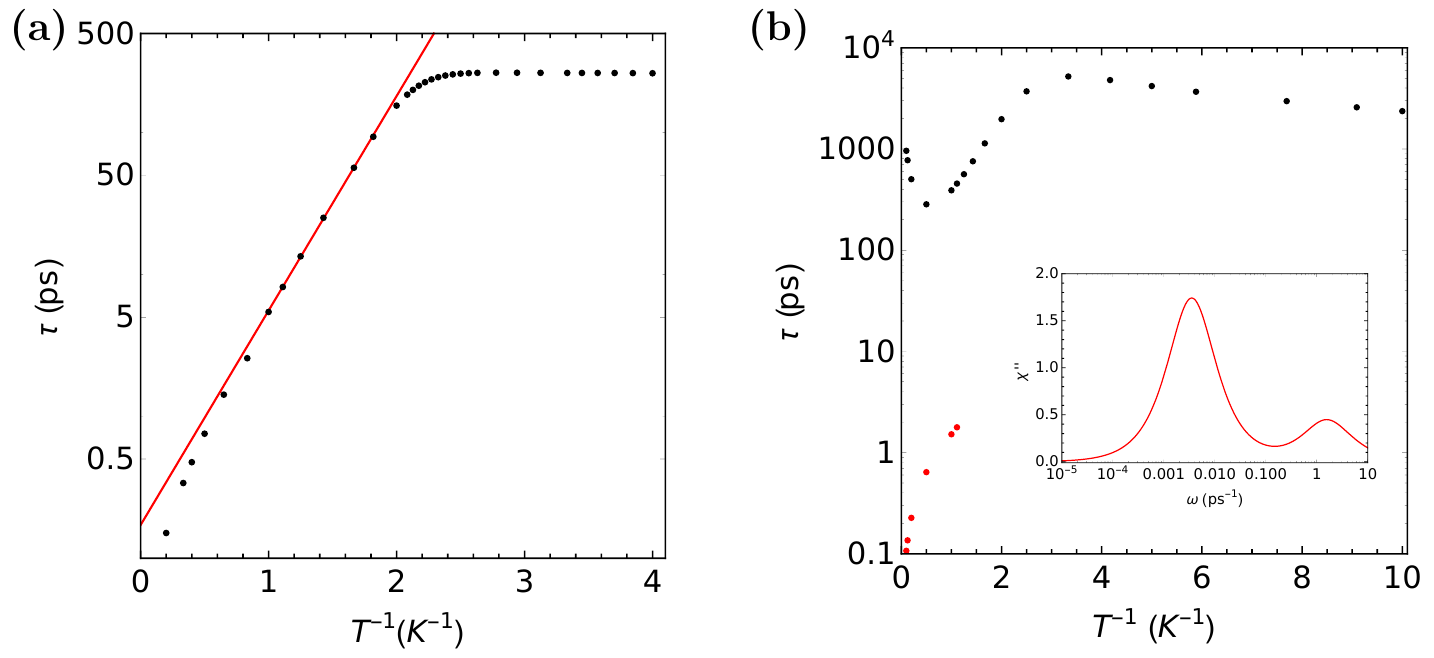}
	\caption{(a) Temperature dependent relaxation times
          of the magnetization in z-direction 
(black dots) for the $S=2$ system with
          $\Operator{X}_{1}$. Fitted Arrhenius law with
          $\tau_0=0.171$~ps and $DS^2=3.48$ (red line).
(b) Temperature dependent dominant relaxation time of the magnetization
in $z$-direction (black dots) for the $S=2$ system with
$\Operator{X}_{2}$. Red dots: second 
relaxation time, which disappears below $T=0.9$~K. Please note the
        different scales for $\tau$.
The inset shows one example of $\chi^{''}$ vs. $\omega$ where
the two peaks belonging to the dominant and second relaxation
times are clearly visible.}  
	\label{jmmm-beckmann-f-3}
\end{figure}

Figure~\ref{jmmm-beckmann-f-3}~(a) shows the calculated relaxation time for the
$S=2$ system with $\op{X}=\op{X}_{1}$. For temperatures between
$0.5\text{ K}$ and $1\text{ K}$ the relaxation times are well described
by the Arrhenius law 
\begin{eqnarray}
\label{E-4-1}
\tau & = & \tau_{0}\cdot\exp\left[\frac{\Delta}{T}\right]
\end{eqnarray}
with fitted parameters $\tau_{0}=0.17144\pm0.008546\text{ ps}$ and
$\Delta=3.48091\pm0.03101\text{ K}$. This is in quite good agreement
with the theoretical value of $\Delta_{\text{th}}=DS^{2}=4\text{ K}$.
Below $0.5\text{ K}$ the resonant tunneling becomes the dominant
process and the relaxation time becomes independent of
temperature \cite{QTM94}. 

Figure~\ref{jmmm-beckmann-f-3}~(b) shows the temperature
dependent relaxation times 
for the $S=2$ system with $\op{X}=\op{X}_{2}$. Here we find for
temperatures larger than $0.9\text{ K}$ a second relaxation time.
Below $0.9\text{ K}$ it is no longer visible because its peak in
$\chi^{''}$ is masked by the peak belonging to the dominant
relaxation time. The dominant relaxation time itself shows an unphysical
behavior: at higher temperatures it is again increasing with
increasing temperature. This is in contradiction to
experimentally measured data as well as to the whole idea of
activated behavior. We verified that this observation is not an
artifact 
of the relatively high value of the coupling constant $\lambda$.
The observed unphysical behavior is also present with much smaller
values of $\lambda$, but smaller values shift the strange behavior
to higher temperatures. We also noticed that the unphysical
behavior is present for any amount of $\op{S}_{z}$ in the
transition operator $\op{X}$. We thus conjecture that the
unphysical behavior is solely due to the presence of
$\op{S}_{z}$ in the transition operator. Its use as a transition
operator thus appears questionable, although other observables
such as the magnetization investigated in section~\ref{sec-3} do
not show (obvious) unphysical behavior.

\begin{figure}[ht!]
	\centering
		\includegraphics[width=0.45\textwidth]{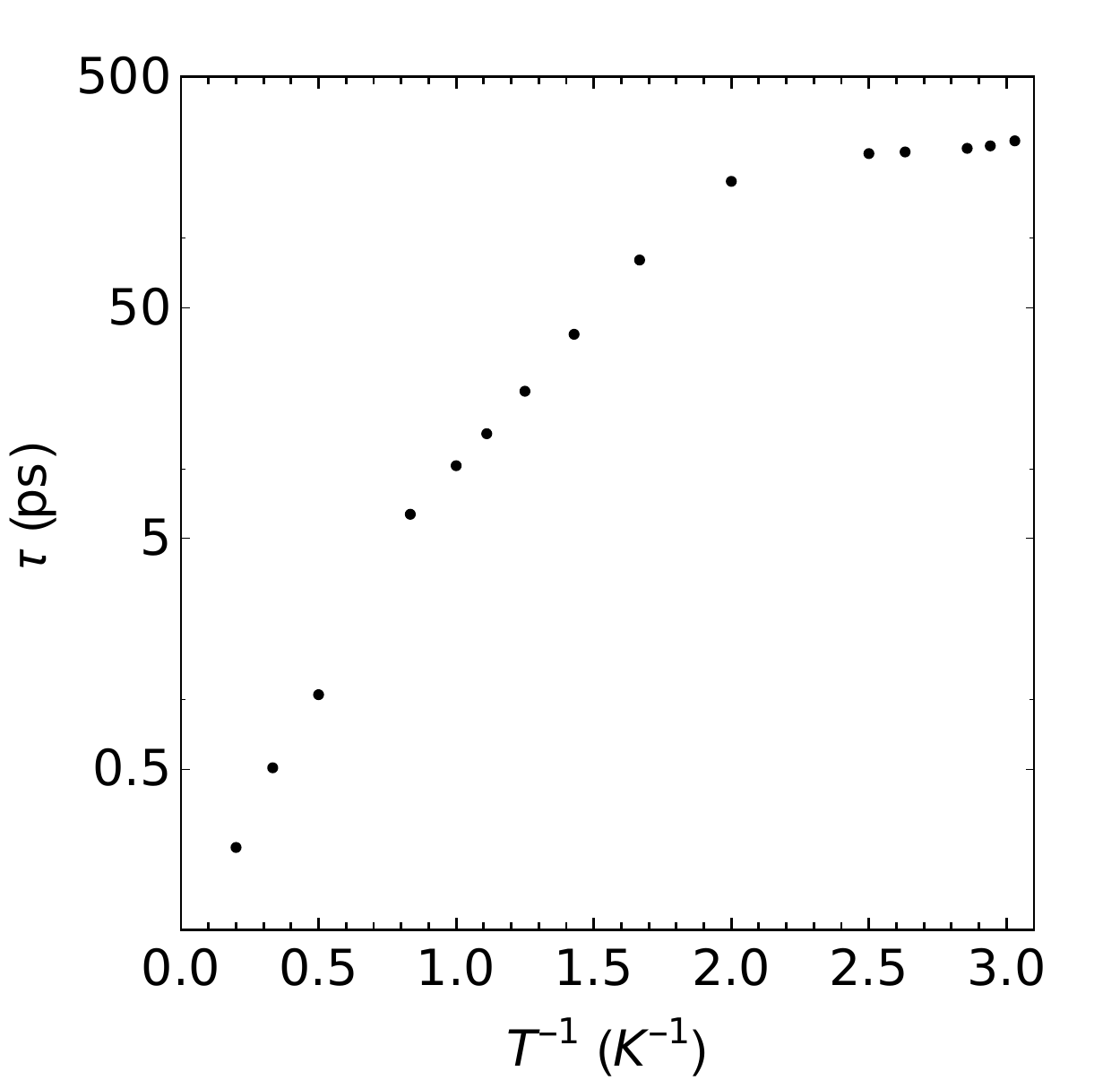}
	\caption{Temperature dependent relaxation time of the magnetization
in $z$-direction (black dots) for the $S=2$ system with $\Operator{X}_{3}$.} 
	\label{jmmm-beckmann-f-4}
\end{figure}

In order to investigate the influence of absorption and emission of
phonons with $\Delta m=\pm 2$ we also calculated the relaxation times
for the $S=2$ system with $\op{X}=\op{X}_{3}$. The results are shown
in figure \ref{jmmm-beckmann-f-4}. Firstly, $\op{X}=\op{X}_{3}$ does not
lead to (obvious) unphysical behavior. Secondly, it is impossible to fit the
curve with one standard Arrhenius law \fmref{E-4-1}. At least two
of them would be necessary, one above and one below $T=1$~K. Qualitatively
the relaxation times in the regime of resonant tunneling, where the
relaxation time becomes independent of the temperature, are nearly
the same for $\op{X}=\op{X}_{3}$ and $\op{X}=\op{X}_{1}$. This is not an effect
of the large amount of $\op{S}_{x}$ in $\op{X}_{3}$. If
$\op{X}=\op{S}_{x}^{2}$ is chosen (not shown here) the same
behavior occurs.

\section{Summary}
\label{sec-5}

In this article we report on numerical studies of the quantum
dynamics of giant spin models by means of the Lindblad scheme. Such
equations of motion for the density matrix allow to treat
unitary as well as relaxation dynamics together with the
prospect of simulating really large spin systems
\cite{ScS:IRPC10,HaS:PRB15} in the future. 

Our concern was
to investigate the influence of the employed transition
operators on non-equilibrium observables such as the
magnetization. We demonstrated that it is possible to evaluate
magnetization processes as well as ac susceptibilities. From the
latter relaxation times can be extracted, and their dependence on
frequency and temperature can be studied. This provides an
additional and very valuable tool to better understand the
experimental ac data.

Unfortunately, the method is limited by two factors. A useful
comparison between theory and experiment needs a detailed
knowledge of the transition operator $\op{X}$, upon which
non-equilibrium observables depend strongly as we showed in this
work. The second limitation is common to all kinds of
time-dependent simulations: the time-step of a numerical
integration is determined by the fastest processes in the
system. This scale is set by the apparent frequencies of
electronic magnetic systems which are in the GHz
range. Therefore, processes that need seconds, minutes or longer
so far cannot be simulated with reasonable effort
\cite{SMD:PRB99}.

In a future investigation we plan to study how transition
operators of the form
\begin{eqnarray}
\label{E-5-1}
\op{X}_{q} 
& = & 
\sum_{\alpha<\beta=x,y,z} \left(\op{S}_{\alpha}\op{S}_{\beta}
+\op{S}_{\beta}\op{S}_{\alpha}\right)/2
\end{eqnarray}
perform in the quantum master equation. Such quadrupolar
operators are supposed to represent the main contribution to the
magnetoelastic coupling between spins and phonons
\cite{SCL:PRL05,CGS:PRL08}.

\section*{Acknowledgment}

The authors thank Ben Balz for useful discussions. Funding by
the Deutsche Forschungsgemeinschaft (DFG SCHN 615/23-1) is thankfully
acknowledged.


\begin{thebibliography}{26}
\providecommand{\natexlab}[1]{#1}
\providecommand{\url}[1]{\texttt{#1}}
\providecommand{\urlprefix}{URL }
\expandafter\ifx\csname urlstyle\endcsname\relax
  \providecommand{\doi}[1]{doi:\discretionary{}{}{}#1}\else
  \providecommand{\doi}[1]{doi:\discretionary{}{}{}\begingroup
  \urlstyle{rm}\url{#1}\endgroup}\fi
\providecommand{\bibinfo}[2]{#2}

\bibitem[{Sessoli et~al.(1993)Sessoli, Gatteschi, Caneschi, and
  Novak}]{SGC:Nat93}
\bibinfo{author}{R.~Sessoli}, \bibinfo{author}{D.~Gatteschi},
  \bibinfo{author}{A.~Caneschi}, \bibinfo{author}{M.~A. Novak},
  \bibinfo{title}{Magnetic bistability in a metal-ion cluster},
  \bibinfo{journal}{Nature} \bibinfo{volume}{365} (\bibinfo{year}{1993})
  \bibinfo{pages}{141--143},
  \urlprefix\url{http://dx.doi.org/10.1038/365141a0}.

\bibitem[{Thomas et~al.(1996)Thomas, Lionti, Ballou, Gatteschi, Sessoli, and
  Barbara}]{TLB:Nature96}
\bibinfo{author}{L.~Thomas}, \bibinfo{author}{F.~Lionti},
  \bibinfo{author}{R.~Ballou}, \bibinfo{author}{D.~Gatteschi},
  \bibinfo{author}{R.~Sessoli}, \bibinfo{author}{B.~Barbara},
  \bibinfo{title}{Macroscopic quantum tunnelling of magnetization in a single
  crystal of nanomagnets}, \bibinfo{journal}{Nature} \bibinfo{volume}{383}
  (\bibinfo{year}{1996}) \bibinfo{pages}{145},
  \urlprefix\url{http://dx.doi.org/10.1038/383145a0}.

\bibitem[{Wernsdorfer and Sessoli(1999)}]{WeS:Science99}
\bibinfo{author}{W.~Wernsdorfer}, \bibinfo{author}{R.~Sessoli},
  \bibinfo{title}{Quantum Phase Interference and Parity Effects in Magnetic
  Molecular Clusters}, \bibinfo{journal}{Science} \bibinfo{volume}{284}
  (\bibinfo{year}{1999}) \bibinfo{pages}{133--135},
  \urlprefix\url{http://www.sciencemag.org/content/284/5411/133.abstract}.

\bibitem[{Chiorescu et~al.(2000)Chiorescu, Wernsdorfer, M\"uller, B\"ogge, and
  Barbara}]{CWM:PRL00}
\bibinfo{author}{I.~Chiorescu}, \bibinfo{author}{W.~Wernsdorfer},
  \bibinfo{author}{A.~M\"uller}, \bibinfo{author}{H.~B\"ogge},
  \bibinfo{author}{B.~Barbara}, \bibinfo{title}{Butterfly hysteresis loop and
  dissipative spin reversal in the $S=1/2$, V$_{15}$ molecular complex},
  \bibinfo{journal}{Phys. Rev. Lett.} \bibinfo{volume}{84}
  (\bibinfo{year}{2000}) \bibinfo{pages}{3454--3457},
  \urlprefix\url{http://link.aps.org/doi/10.1103/PhysRevLett.84.3454}.

\bibitem[{Chudnovsky et~al.(2005)Chudnovsky, Garanin, and
  Schilling}]{CGS:PRB05}
\bibinfo{author}{E.~M. Chudnovsky}, \bibinfo{author}{D.~A. Garanin},
  \bibinfo{author}{R.~Schilling}, \bibinfo{title}{Universal mechanism of spin
  relaxation in solids}, \bibinfo{journal}{Phys. Rev. B} \bibinfo{volume}{72}
  (\bibinfo{year}{2005}) \bibinfo{pages}{094426},
  \urlprefix\url{http://link.aps.org/abstract/PRB/v72/e094426}.

\bibitem[{Santini et~al.(2005)Santini, Carretta, Liviotti, Amoretti, Carretta,
  Filibian, Lascialfari, and Micotti}]{SCL:PRL05}
\bibinfo{author}{P.~Santini}, \bibinfo{author}{S.~Carretta},
  \bibinfo{author}{E.~Liviotti}, \bibinfo{author}{G.~Amoretti},
  \bibinfo{author}{P.~Carretta}, \bibinfo{author}{M.~Filibian},
  \bibinfo{author}{A.~Lascialfari}, \bibinfo{author}{E.~Micotti},
  \bibinfo{title}{NMR as a probe of the relaxation of the magnetization in
  magnetic molecules}, \bibinfo{journal}{Phys. Rev. Lett.} \bibinfo{volume}{94}
  (\bibinfo{year}{2005}) \bibinfo{pages}{077203},
  \urlprefix\url{http://link.aps.org/doi/10.1103/PhysRevLett.94.077203}.

\bibitem[{Carretta et~al.(2008)Carretta, Guidi, Santini, Amoretti, Pieper,
  Lake, van Slageren, Hallak, Wernsdorfer, Mutka, Russina, Milios, and
  Brechin}]{CGS:PRL08}
\bibinfo{author}{S.~Carretta}, \bibinfo{author}{T.~Guidi},
  \bibinfo{author}{P.~Santini}, \bibinfo{author}{G.~Amoretti},
  \bibinfo{author}{O.~Pieper}, \bibinfo{author}{B.~Lake},
  \bibinfo{author}{J.~van Slageren}, \bibinfo{author}{F.~E. Hallak},
  \bibinfo{author}{W.~Wernsdorfer}, \bibinfo{author}{H.~Mutka},
  \bibinfo{author}{M.~Russina}, \bibinfo{author}{C.~J. Milios},
  \bibinfo{author}{E.~K. Brechin}, \bibinfo{title}{Breakdown of the Giant Spin
  Model in the Magnetic Relaxation of the Mn[sub 6] Nanomagnets},
  \bibinfo{journal}{Phys. Rev. Lett.} \bibinfo{volume}{100}
  (\bibinfo{year}{2008}) \bibinfo{pages}{157203},
  \urlprefix\url{http://link.aps.org/abstract/PRL/v100/e157203}.

\bibitem[{Garlatti et~al.(2016)Garlatti, Bordignon, Carretta, Allodi, Amoretti,
  De~Renzi, Lascialfari, Furukawa, Timco, Woolfson, Winpenny, and
  Santini}]{GBC:PRB16}
\bibinfo{author}{E.~Garlatti}, \bibinfo{author}{S.~Bordignon},
  \bibinfo{author}{S.~Carretta}, \bibinfo{author}{G.~Allodi},
  \bibinfo{author}{G.~Amoretti}, \bibinfo{author}{R.~De~Renzi},
  \bibinfo{author}{A.~Lascialfari}, \bibinfo{author}{Y.~Furukawa},
  \bibinfo{author}{G.~A. Timco}, \bibinfo{author}{R.~Woolfson},
  \bibinfo{author}{R.~E.~P. Winpenny}, \bibinfo{author}{P.~Santini},
  \bibinfo{title}{Relaxation dynamics in the frustrated ${\mathrm{Cr}}_{9}$
  antiferromagnetic ring probed by NMR}, \bibinfo{journal}{Phys. Rev. B}
  \bibinfo{volume}{93} (\bibinfo{year}{2016}) \bibinfo{pages}{024424},
  \urlprefix\url{http://link.aps.org/doi/10.1103/PhysRevB.93.024424}.

\bibitem[{Lindblad(1976)}]{Lin:CMP76}
\bibinfo{author}{G.~Lindblad}, \bibinfo{title}{On the generators of quantum
  dynamical semigroups}, \bibinfo{journal}{Comm. Math. Phys.}
  \bibinfo{volume}{48} (\bibinfo{year}{1976}) \bibinfo{pages}{119--130},
  \urlprefix\url{http://projecteuclid.org/euclid.cmp/1103899849}.

\bibitem[{Raedt et~al.(1997)Raedt, Miyashita, Saito, García-Pablos, and
  García}]{RMS:PRB97}
\bibinfo{author}{H.~D. Raedt}, \bibinfo{author}{S.~Miyashita},
  \bibinfo{author}{K.~Saito}, \bibinfo{author}{D.~García-Pablos},
  \bibinfo{author}{N.~García}, \bibinfo{title}{Theory of quantum tunneling of
  the magnetization in magnetic particles}, \bibinfo{journal}{Phys. Rev. B}
  \bibinfo{volume}{56} (\bibinfo{year}{1997}) \bibinfo{pages}{11761},
  \urlprefix\url{http://link.aps.org/doi/10.1103/PhysRevB.56.11761}.

\bibitem[{Thorwart et~al.(1998{\natexlab{a}})Thorwart, Reimann, Jung, and
  Fox}]{TRJ:PLA98}
\bibinfo{author}{M.~Thorwart}, \bibinfo{author}{P.~Reimann},
  \bibinfo{author}{P.~Jung}, \bibinfo{author}{R.~Fox}, \bibinfo{title}{Quantum
  steps in hysteresis loops}, \bibinfo{journal}{Phys. Lett. A}
  \bibinfo{volume}{239} (\bibinfo{year}{1998}{\natexlab{a}})
  \bibinfo{pages}{233--238},
  \urlprefix\url{http://www.sciencedirect.com/science/article/pii/S0375960198000218}.

\bibitem[{Miyashita et~al.(1998)Miyashita, Saito, and Raedt}]{MSD:PRL98}
\bibinfo{author}{S.~Miyashita}, \bibinfo{author}{K.~Saito},
  \bibinfo{author}{H.~D. Raedt}, \bibinfo{title}{Nontrivial Response of
  Nanoscale Uniaxial Magnets to an Alternating Field}, \bibinfo{journal}{Phys.
  Rev. Lett.} \bibinfo{volume}{80} (\bibinfo{year}{1998})
  \bibinfo{pages}{1525},
  \urlprefix\url{http://link.aps.org/doi/10.1103/PhysRevLett.80.1525}.

\bibitem[{Saito et~al.(1999)Saito, Miyashita, and De~Raedt}]{SMD:PRB99}
\bibinfo{author}{K.~Saito}, \bibinfo{author}{S.~Miyashita},
  \bibinfo{author}{H.~De~Raedt}, \bibinfo{title}{Effects of the environment on
  nonadiabatic magnetization process in uniaxial molecular magnets at very low
  temperatures}, \bibinfo{journal}{Phys. Rev. B} \bibinfo{volume}{60}
  (\bibinfo{year}{1999}) \bibinfo{pages}{14553--14556},
  \urlprefix\url{http://link.aps.org/doi/10.1103/PhysRevB.60.14553}.

\bibitem[{Saito and Miyashita(2001)}]{SaM:JPSJ01}
\bibinfo{author}{K.~Saito}, \bibinfo{author}{S.~Miyashita},
  \bibinfo{title}{Magnetic Foehn Effect in Adiabatic Transition},
  \bibinfo{journal}{J. Phys. Soc. Jpn.} \bibinfo{volume}{70}
  (\bibinfo{year}{2001}) \bibinfo{pages}{3385--3390},
  \urlprefix\url{http://dx.doi.org/10.1143/JPSJ.70.3385}.

\bibitem[{Nakano and Miyashita(2001)}]{NaM:JPSJ01}
\bibinfo{author}{H.~Nakano}, \bibinfo{author}{S.~Miyashita},
  \bibinfo{title}{Magnetization Process of Nanoscale Iron Cluster},
  \bibinfo{journal}{J. Phys. Soc. Jpn.} \bibinfo{volume}{70}
  (\bibinfo{year}{2001}) \bibinfo{pages}{2151--2157},
  \urlprefix\url{http://dx.doi.org/10.1143/JPSJ.70.2151}.

\bibitem[{Kawakami et~al.(2009)Kawakami, Nitta, Takahata, Shoji, Kitagawa,
  Nakano, Okumura, and Yamaguchi}]{KNT:P09}
\bibinfo{author}{T.~Kawakami}, \bibinfo{author}{H.~Nitta},
  \bibinfo{author}{M.~Takahata}, \bibinfo{author}{M.~Shoji},
  \bibinfo{author}{Y.~Kitagawa}, \bibinfo{author}{M.~Nakano},
  \bibinfo{author}{M.~Okumura}, \bibinfo{author}{K.~Yamaguchi},
  \bibinfo{title}{Quantum dynamic simulations for single molecular magnets
  using anisotropic spin models}, \bibinfo{journal}{Polyhedron}
  \bibinfo{volume}{28} (\bibinfo{year}{2009}) \bibinfo{pages}{2092--2096},
  \urlprefix\url{http://dx.doi.org/10.1016/j.poly.2009.02.032}.

\bibitem[{Chiesa et~al.(2014)Chiesa, Gerace, Troiani, Amoretti, Santini, and
  Carretta}]{CGT:PRA14}
\bibinfo{author}{A.~Chiesa}, \bibinfo{author}{D.~Gerace},
  \bibinfo{author}{F.~Troiani}, \bibinfo{author}{G.~Amoretti},
  \bibinfo{author}{P.~Santini}, \bibinfo{author}{S.~Carretta},
  \bibinfo{title}{Robustness of quantum gates with hybrid spin-photon qubits in
  superconducting resonators}, \bibinfo{journal}{Phys. Rev. A}
  \bibinfo{volume}{89} (\bibinfo{year}{2014}) \bibinfo{pages}{052308},
  \urlprefix\url{http://link.aps.org/doi/10.1103/PhysRevA.89.052308}.

\bibitem[{Thorwart et~al.(1998{\natexlab{b}})Thorwart, Reimann, Jung, and
  Fox}]{TRJ:CP98}
\bibinfo{author}{M.~Thorwart}, \bibinfo{author}{P.~Reimann},
  \bibinfo{author}{P.~Jung}, \bibinfo{author}{R.~Fox}, \bibinfo{title}{Quantum
  hysteresis and resonant tunneling in bistable systems},
  \bibinfo{journal}{Chem. Phys.} \bibinfo{volume}{235}
  (\bibinfo{year}{1998}{\natexlab{b}}) \bibinfo{pages}{61 -- 80},
  \urlprefix\url{http://www.sciencedirect.com/science/article/pii/S0301010498001281}.

\bibitem[{Liddle and van Slageren(2015)}]{LvS:CSR15}
\bibinfo{author}{S.~T. Liddle}, \bibinfo{author}{J.~van Slageren},
  \bibinfo{title}{Improving f-element single molecule magnets},
  \bibinfo{journal}{Chem. Soc. Rev.} \bibinfo{volume}{44}
  (\bibinfo{year}{2015}) \bibinfo{pages}{6655--6669},
  \urlprefix\url{http://dx.doi.org/10.1039/C5CS00222B}.

\bibitem[{Cervetti et~al.(2016)Cervetti, Rettori, Pini, Cornia, Repolles, Luis,
  Dressel, Rauschenbach, Kern, Burghard, and Bogani}]{CRP:NM16}
\bibinfo{author}{C.~Cervetti}, \bibinfo{author}{A.~Rettori},
  \bibinfo{author}{M.~G. Pini}, \bibinfo{author}{A.~Cornia},
  \bibinfo{author}{A.~Repolles}, \bibinfo{author}{F.~Luis},
  \bibinfo{author}{M.~Dressel}, \bibinfo{author}{S.~Rauschenbach},
  \bibinfo{author}{K.~Kern}, \bibinfo{author}{M.~Burghard},
  \bibinfo{author}{L.~Bogani}, \bibinfo{title}{The classical and quantum
  dynamics of molecular spins on graphene}, \bibinfo{journal}{Nat. Mater.}
  \bibinfo{volume}{15} (\bibinfo{year}{2016}) \bibinfo{pages}{164--168},
  \urlprefix\url{http://dx.doi.org/10.1038/nmat4490}.

\bibitem[{Caneschi et~al.(1999)Caneschi, Gatteschi, Sangregorio, Sessoli,
  Sorace, Cornia, Novak, Paulsen, and Wernsdorfer}]{CGS:JMMM99}
\bibinfo{author}{A.~Caneschi}, \bibinfo{author}{D.~Gatteschi},
  \bibinfo{author}{C.~Sangregorio}, \bibinfo{author}{R.~Sessoli},
  \bibinfo{author}{L.~Sorace}, \bibinfo{author}{A.~Cornia},
  \bibinfo{author}{M.~A. Novak}, \bibinfo{author}{C.~Paulsen},
  \bibinfo{author}{W.~Wernsdorfer}, \bibinfo{title}{The molecular approach to
  nanoscale magnetism}, \bibinfo{journal}{J. Magn. Magn. Mater.}
  \bibinfo{volume}{200} (\bibinfo{year}{1999}) \bibinfo{pages}{182 -- 201},
  \urlprefix\url{http://www.sciencedirect.com/science/article/pii/S0304885399004084}.

\bibitem[{Prokof'ev and Stamp(2000)}]{PrS:RPP00}
\bibinfo{author}{N.~V. Prokof'ev}, \bibinfo{author}{P.~C.~E. Stamp},
  \bibinfo{title}{Theory of the spin bath}, \bibinfo{journal}{Rep. Prog. Phys.}
  \bibinfo{volume}{63} (\bibinfo{year}{2000}) \bibinfo{pages}{669--726},
  \urlprefix\url{http://iopscience.iop.org/article/10.1088/0034-4885/63/4/204}.

\bibitem[{Gatteschi et~al.(2006)Gatteschi, Sessoli, and Villain}]{GSV:2006}
\bibinfo{author}{D.~Gatteschi}, \bibinfo{author}{R.~Sessoli},
  \bibinfo{author}{J.~Villain}, \bibinfo{title}{Molecular Nanomagnets},
  Mesoscopic Physics and Nanotechnology, \bibinfo{publisher}{Oxford University
  Press}, \bibinfo{address}{Oxford}, \bibinfo{year}{2006}.

\bibitem[{Gunther and Barbara(1995)}]{QTM94}
\bibinfo{editor}{L.~Gunther}, \bibinfo{editor}{B.~Barbara} (Eds.),
  \bibinfo{title}{Quantum Tunneling of Magnetization — QTM '94}, vol.
  \bibinfo{volume}{301} of \emph{\bibinfo{series}{NATO ASI Series}},
  \bibinfo{publisher}{Springer}, \bibinfo{year}{1995}.

\bibitem[{Schnalle and Schnack(2010)}]{ScS:IRPC10}
\bibinfo{author}{R.~Schnalle}, \bibinfo{author}{J.~Schnack},
  \bibinfo{title}{Calculating the energy spectra of magnetic molecules:
  application of real- and spin-space symmetries}, \bibinfo{journal}{Int. Rev.
  Phys. Chem.} \bibinfo{volume}{29} (\bibinfo{year}{2010})
  \bibinfo{pages}{403--452},
  \urlprefix\url{http://dx.doi.org/10.1080/0144235X.2010.485755}.

\bibitem[{Hanebaum and Schnack(2015)}]{HaS:PRB15}
\bibinfo{author}{O.~Hanebaum}, \bibinfo{author}{J.~Schnack},
  \bibinfo{title}{Thermodynamic observables of ${\mathrm{Mn}}_{12}$-acetate
  calculated for the full spin Hamiltonian}, \bibinfo{journal}{Phys. Rev. B}
  \bibinfo{volume}{92} (\bibinfo{year}{2015}) \bibinfo{pages}{064424},
  \urlprefix\url{http://link.aps.org/doi/10.1103/PhysRevB.92.064424}.

\end{thebibliography}

\end{document}